\documentclass[prl,reprint,notitlepage,nofootinbib]{revtex4-1}

\usepackage{graphicx}
\usepackage{amsmath}
\usepackage{amsfonts}
\usepackage[utf8]{inputenc}
\usepackage{slashed}
\usepackage{hyperref}
\hypersetup{colorlinks, linkcolor = [rgb]{0,0.0,0.75}, citecolor = [rgb]{0,0.0,0.75}, urlcolor = [rgb]{0,0.0,0.75}}

\DeclareMathOperator{\Tr}{Tr}
\DeclareMathOperator{\sgn}{sgn}
\DeclareMathOperator{\Ci}{Ci}
\newcommand{\cO}{\mathcal{O}}
\newcommand{\MSbar}{\overline{\text{MS}}}

\begin{document}

\title{Nonperturbative renormalization of nonlocal quark bilinears for quasi-PDFs on the lattice using an auxiliary field}

\author{Jeremy~Green}
\email{jeremy.green@desy.de}
\author{Karl~Jansen}
\author{Fernanda~Steffens}
\affiliation{NIC, Deutsches Elektronen-Synchrotron, 15738 Zeuthen, Germany}

\date{\today}

\begin{abstract}
  Quasi-PDFs provide a path toward an \emph{ab initio} calculation of
  parton distribution functions (PDFs) using lattice QCD. One of the
  problems faced in calculations of quasi-PDFs is the renormalization
  of a nonlocal operator. By introducing an auxiliary field, we can
  replace the nonlocal operator with a pair of local operators in an
  extended theory. On the lattice, this is closely related to the
  static quark theory. In this approach, we show how to understand the
  pattern of mixing that is allowed by chiral symmetry breaking, and
  obtain a master formula for renormalizing the nonlocal operator that
  depends on three parameters. We present an approach for
  nonperturbatively determining these parameters and use perturbation
  theory to convert to the $\MSbar$ scheme. Renormalization parameters
  are obtained for two lattice spacings using Wilson twisted mass
  fermions and for different discretizations of the Wilson line in the
  nonlocal operator. Using these parameters we show the effect of
  renormalization on nucleon matrix elements with pion mass
  approximately 370~MeV, and compare renormalized results for the two
  lattice spacings. The renormalized matrix elements are consistent
  among the different Wilson line discretizations and lattice
  spacings.
\end{abstract}

\maketitle

Parton distribution functions (PDFs) describe the distribution of
quarks and gluons inside a proton with respect to its longitudinal
momentum. They are universal: the same PDFs appear in many different
scattering processes, and they are phenomenologically determined from
global fits to collider data. Except for their lowest Mellin moments,
PDFs have resisted \emph{ab initio} calculation. Lattice QCD can be
used to calculate many properties of protons, but it is an inherently
Euclidean space method, whereas PDFs are defined in Minkowski space
via the matrix elements of operators with quark\footnote{Here we focus
  on quark and antiquark PDFs rather than gluon PDFs.} creation and
annihilation separated along the light cone. A possible solution to
this problem was proposed by Ji~\cite{Ji:2013dva}: compute
\emph{quasi-PDFs} using matrix elements of equal-time operators
\begin{equation}
  \cO_\Gamma(x,\xi,n) \equiv
  \bar\psi(x+\xi n) \Gamma W(x+\xi n,x)\psi(x),
\end{equation}
where the $\psi$ and $\bar\psi$ are spatially separated by distance
$\xi$ in direction $n$ and connected by a straight Wilson line
$W$. From quasi-PDFs we can obtain PDFs via a matching
formula~\cite{Xiong:2013bka,Alexandrou:2015rja}, in the limit where
the proton's momentum component $p\cdot n$ goes to infinity.

One of the challenges in lattice calculations of quasi-PDFs is
renormalization of $\cO_\Gamma$, which is a nonlocal operator and is
known to be power-law divergent~\cite{Ishikawa:2016znu,Chen:2016fxx}.
This operator also appears in the related approach of
pseudo-PDFs~\cite{Radyushkin:2017cyf,Orginos:2017kos}, and similar
nonlocal operators are used to study transverse momentum-dependent
PDFs~\cite{Musch:2011er,Engelhardt:2015xja,Yoon:2017qzo}.
The initial lattice studies of
quasi-PDFs~\cite{Lin:2014zya,Alexandrou:2015rja,Chen:2016utp,Alexandrou:2016jqi}
did not include a complete renormalization, but they did build the
Wilson line using smeared gauge links, which has been shown in
perturbation theory to reduce the power
divergence~\cite{Ishikawa:2016znu}.

Renormalization of $\cO_\Gamma$ was studied in one-loop lattice
perturbation theory in Ref.~\cite{Constantinou:2017sej}, where it was
found that chiral symmetry breaking allows $\cO_\Gamma$ to mix with
$\cO_{\{\slashed{n},\Gamma\}}$. Numerical evidence for this mixing was
also found in Ref.~\cite{Yoon:2017qzo}.
The study of nonperturbative renormalization was pioneered in
Refs.~\cite{Alexandrou:2017huk,Chen:2017mzz}, which used the
Rome-Southampton method~\cite{Martinelli:1994ty} to obtain a complex
$\xi$-dependent renormalization factor (or matrix, when there is
mixing) $Z(\xi)$ in the RI-MOM scheme. This was then supplemented by a
perturbative conversion~\cite{Constantinou:2017sej}
to the $\MSbar$ scheme used in
phenomenology. The problems with this method are that a whole
function, rather than a handful of parameters, must be determined, and
that conversion at large $\xi$ may occur outside the regime where
perturbation theory is valid. Since the intermediate scheme fixes more
than the minimal number of parameters, this means that nonperturbative
information in correlation functions is lost, only to be recovered
perturbatively in the conversion to $\MSbar$.

In this work, we study the use of an auxiliary scalar, color triplet
field $\zeta(\xi)$ defined only on the line $x+\xi n$ to simplify the
renormalization of $\cO_\Gamma$. In this approach, we replace
correlation functions in QCD involving $\cO_\Gamma$ with correlation
functions in the extended theory QCD+$\zeta$ involving the local color
singlet bilinear $\phi\equiv \bar\zeta\psi$. This approach has been
used long ago in the continuum~\cite{Craigie:1980qs,Dorn:1986dt}.

On the lattice, for now we restrict $n$ to point along one of the
axes, $n=\pm\hat\mu$, and use the action
\begin{equation}
\begin{gathered}
  S_\zeta = a\sum_\xi\frac{1}{1+am_0}\bar\zeta(x+\xi n)\left[\nabla_n+m_0\right]\zeta(x+\xi n),\\
\nabla_n\equiv\begin{cases}
n\cdot\nabla^* = \nabla^*_\mu & \text{if }n=+\hat\mu \\
n\cdot\nabla = -\nabla_\mu & \text{if }n=-\hat\mu,
\end{cases}
\end{gathered}
\end{equation}
where $\nabla$ and $\nabla^*$ are the forward and backward lattice
covariant derivatives and $a$ is the lattice spacing. This yields the
bare propagator in fixed gauge background
\begin{equation}
  \left\langle \zeta(x+\xi n)\bar\zeta(x)\right\rangle_\zeta = \theta(\xi)e^{-m\xi}W(x+\xi n,x),
\end{equation}
where $m=a^{-1}\log(1+am_0)$ and $W$ is the simple product of lattice
gauge links connecting $x$ and $x+\xi n$. Smeared gauge links can be
used to construct $W$ by using the same gauge links to define
$\nabla_n$. The mass term can not be forbidden by any symmetry, and
corresponds to an $O(a^{-1})$ counterterm. Using this propagator, we
obtain for $m=0$ and $\xi>0$ the relation
\begin{equation}
  \cO_\Gamma(x,\xi,n) = \left\langle \bar\phi(x+\xi n)\Gamma \phi(x) \right\rangle_\zeta.
\end{equation}
To obtain $\cO_\Gamma$ for $\xi<0$, we reverse the direction in which
$\zeta$ propagates, and use
$\cO_\Gamma(x,\xi,n)=\cO_\Gamma(x,-\xi,-n)$.

In addition to determining the counterterm $m_0$, we must also
renormalize the local composite operator $\phi$. The lattice quark
action may break chiral symmetry, in which case $\phi$ can mix with
$\slashed{n}\phi$. We thus obtain the renormalization pattern
\begin{equation}
  \phi_R = Z_\phi\left(\phi+r_\text{mix}\slashed{n}\phi\right),\quad
  \bar\phi_R = Z_\phi\left(\bar\phi+r_\text{mix}\bar\phi\slashed{n}\right).
\end{equation}
We can also use projectors to form operators
$\phi^\pm\equiv\frac{1}{2}(1\pm\slashed{n})\phi$ that renormalize
diagonally with $Z_\phi^\pm\equiv Z_\phi(1\pm r_\text{mix})$. This pattern
leads to the form of the renormalized $\cO_\Gamma$, for $\xi\neq 0$:
\begin{equation}
\begin{gathered}
  \cO_\Gamma^R(x,\xi,n) = Z_\phi^2e^{-m|\xi|}\cO_{\Gamma'}(x,\xi,n),\\
\Gamma' = \Gamma + r_\text{mix}\sgn(\xi)\{\slashed{n},\Gamma\} + r_\text{mix}^2\slashed{n}\Gamma\slashed{n}.
\end{gathered}
\end{equation}
$\cO_\Gamma$ can therefore be renormalized by determining three
parameters: the linearly divergent $m$, the log-divergent $Z_\phi$,
and the finite $r_\text{mix}$. Since $r_\text{mix}$ is $O(g^2)$, this
is the same pattern at one-loop order as in
Ref.~\cite{Constantinou:2017sej}. At $\xi=0$, $\cO_\Gamma$ is a local
quark bilinear with different divergence structure, and should have a
separate renormalization factor~\cite{Constantinou_privcomm}, which
can be computed in the usual way; we use results from
Ref.~\cite{Alexandrou:2015sea}. We also note that since the local
operator $\phi$ is not flavor singlet, there is no mixing between
quark and gluon quasi-PDFs even when $\cO_\Gamma$ is flavor singlet.
In the latter case mixing will occur in the matching to PDFs.

If we choose $n=\hat t$ and give $\zeta$ spin degrees of freedom
(which do not couple in the action), then $S_\zeta$ becomes the action
for a static quark on the
lattice~\cite{Eichten:1989zv,Sommer:2010ic}. In particular, $\phi$ is
related to the static-light bilinears and we find relations between
renormalization factors: $Z_V^\text{stat}=Z_\phi^+$ and
$Z_A^\text{stat}=Z_\phi^-$. The static quark theory also tells us how
to remove $O(a)$ lattice artifacts~\cite{Kurth:2000ki}, which are
present even if chiral symmetry is preserved on the lattice. In the
continuum, the relation between $\cO_\Gamma$ and the static quark
theory was previously discussed in Ref.~\cite{Ji:2015jwa}.

We determine the renormalization parameters nonperturbatively using a
variant of the Rome-Southampton method. In Landau gauge on $N_f=4$
twisted mass lattice ensembles~\cite{Alexandrou:2015sea}, we compute
the position-space $\zeta$ propagator,
\begin{equation}
  S_\zeta(\xi) \equiv \left\langle \zeta(x+\xi n)\bar\zeta(x)\right\rangle_{\text{QCD}+\zeta}
  = \bigl\langle W(x+\xi n,x)\bigr\rangle_\text{QCD},
\end{equation}
the momentum-space quark propagator $S_\psi(p)$, and the mixed-space
Green's function for $\phi^\pm$:
\begin{equation}
  G^\pm(\xi,p) \equiv
  \int d^4x\, e^{ip\cdot x}\left\langle
  \zeta(\xi n)\phi^{\pm}(0)\bar\psi(x)\right\rangle_{\text{QCD}+\zeta}.
\end{equation}
The renormalization parameters, as well as the $\zeta$ and $\psi$
field renormalizations can be determined by the conditions
\begin{gather}
  -\frac{d}{d\xi}\log\Tr S_\zeta(\xi)\Bigr|_{\xi=\xi_0} + m=0,\label{eq:cond_m}\\
\left[\frac{Z_\zeta}{3}\Tr S_\zeta(\xi_0)\right]^2=\frac{Z_\zeta}{3}\Tr S_\zeta(2\xi_0),\label{eq:cond_zeta}\\
\frac{1}{6}\frac{Z_\phi^\pm}{\sqrt{Z_\zeta Z_\psi}}\Re\Tr\left[S_\zeta^{-1}(\xi_0)G^\pm(\xi_0,p_0)S_\psi^{-1}(p_0)\right]=1,\label{eq:cond_phi}
\end{gather}
and the RI$'$-MOM/RI-SMOM condition for
$S_\psi$~\cite{Martinelli:1994ty,Sturm:2009kb}. Eq.~\eqref{eq:cond_m}
is sensitive
to $m$, whereas the others are constructed to eliminate dependence on
it.  These conditions define a two-parameter family of renormalization
schemes at scale $\mu^2=p_0^2$, that depend on the dimensionless
quantities $y\equiv|p_0|\xi_0$ and $z\equiv p_0\cdot
n/|p_0|$. We call this family of schemes RI-xMOM.
Restricting to the kinematics $p_0\propto n$ (i.e.\ $z=1$),
we have computed the conversion to the $\MSbar$ scheme at one-loop
order, using dimensionally regularized perturbation
theory~\cite{Green_forthcoming}. We obtain in Landau gauge:
\begin{equation}
\begin{aligned}\label{eq:conversion}
   C_\phi &\equiv \frac{Z_\phi^{\MSbar}(\mu)}{Z_\phi^\text{RI-xMOM}(\mu,y,z=1)} \\
&= 1
+ \frac{\alpha_s C_F}{8\pi}\biggl[
  6\log\frac{y}{4}
  + 6\gamma_E - 8\log 2 + 7 \\
&\qquad\qquad - \cos y - \left(8\cos\frac{y}{2} - y\sin\frac{y}{2}\right)
\Ci\left(\frac{y}{2}\right) \\
&\qquad\qquad + 8\Ci(y)\biggr] + O(\alpha_s^2),
\end{aligned}
\end{equation}
where $\gamma_E$ is the Euler-Mascheroni constant and $\Ci$ is the
cosine integral function, $\Ci(z) \equiv -\int_z^\infty
\frac{\cos(t)}{t} dt$. For converting $m$ to the $\MSbar$ scheme, we
use the three-loop results for the static quark propagator from
Refs.~\cite{Chetyrkin:2003vi,Melnikov:2000zc}.

\begin{figure}
  \centering
  \includegraphics[width=\columnwidth]{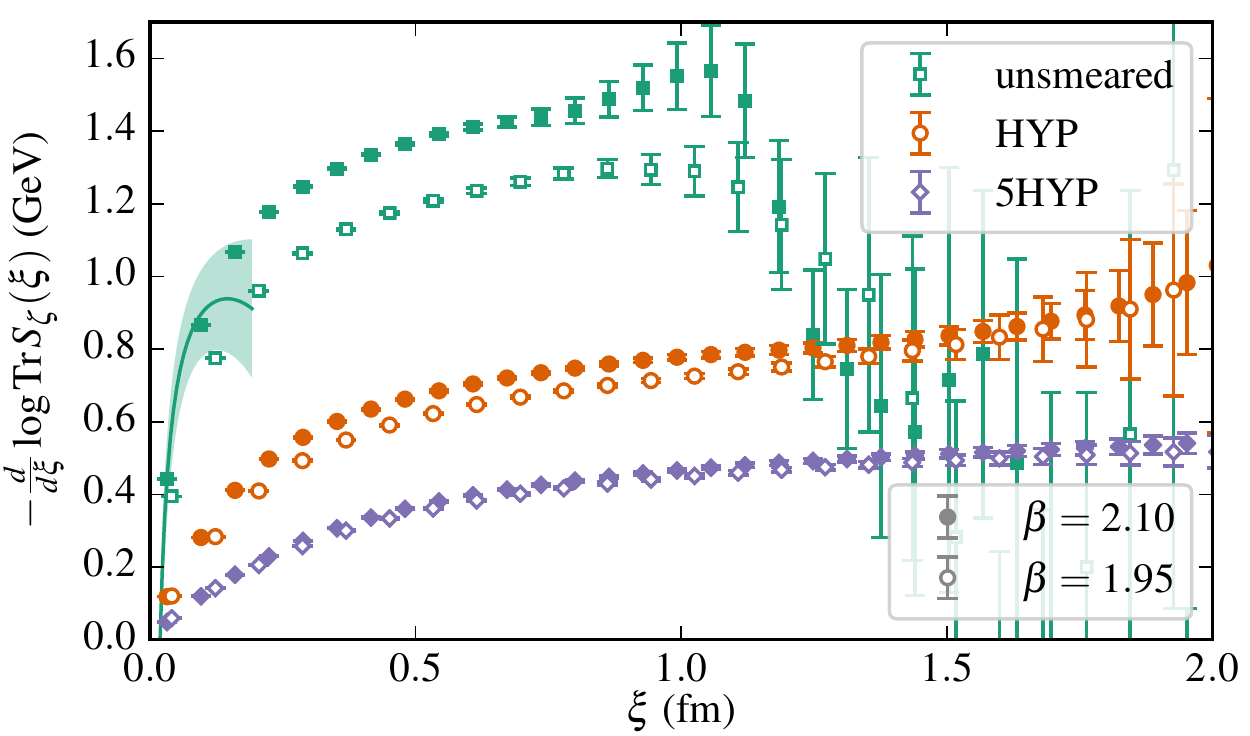}
  \caption{Effective energy of the bare auxiliary field propagator, for two
    lattice spacings and three different link discretizations. Solid
    symbols show the finer lattice spacing and open symbols show the
    coarser one. The curve shows the three-loop perturbative result,
    shifted vertically by $-m$ to match it to the unsmeared data on the
    finer lattice spacing. Its error band indicates the size of the
    $O(\alpha_s^3)$ contribution.}
  \label{fig:Eeff}
\end{figure}

In Fig.~\ref{fig:Eeff}, we show the quantity\footnote{This is the
  ``effective energy'' of the auxiliary field propagator. It was
  previously studied in Ref.~\cite{Musch:2010ka}, where it was denoted
  $Y_\text{line}$.} in Eq.~\eqref{eq:cond_m}, for two different
lattice spacings: $a=0.082$~fm ($\beta=1.95$) and $a=0.064$~fm
($\beta=2.10$). This is renormalized by adding $m$. Without smearing
there is a significant difference between the two lattice spacings due
to the linear divergence, but this is greatly reduced by applying one
or five steps of HYP smearing~\cite{Hasenfratz:2001hp}, which also
reduces the statistical uncertainty at large $\xi$. At small $\xi/a$,
smearing distorts the shape and therefore we choose to impose our
condition at $\xi_0\approx 0.6$~fm, where the shapes are similar. We
then convert to the $\MSbar$ scheme at short distance by matching the
result using unsmeared links on the fine ensemble to the perturbative
result.

From Eq.~\eqref{eq:cond_phi} an estimator for $r_\text{mix}$ can be
isolated. In our data we see indications that this suffers from
significant $O(a)$ lattice artifacts. These could be remedied through
Symanzik improvement~\cite{Kurth:2000ki}, but instead we choose to
focus on the helicity quasi-PDF, which is unaffected by
mixing\footnote{Following the same logic used to derive automatic
$O(a)$ improvement~\cite{Frezzotti:2003ni}, it can also be argued that for
twisted mass lattice QCD at maximal twist, the contribution from the
mixing operator to nucleon matrix elements will vanish at $O(a)$.} and
depends only on $r_\text{mix}^2$. We find that the latter is not much
greater than 1\% and can be neglected at the current level of
precision.

\begin{figure}
  \centering
  \includegraphics[width=\columnwidth]{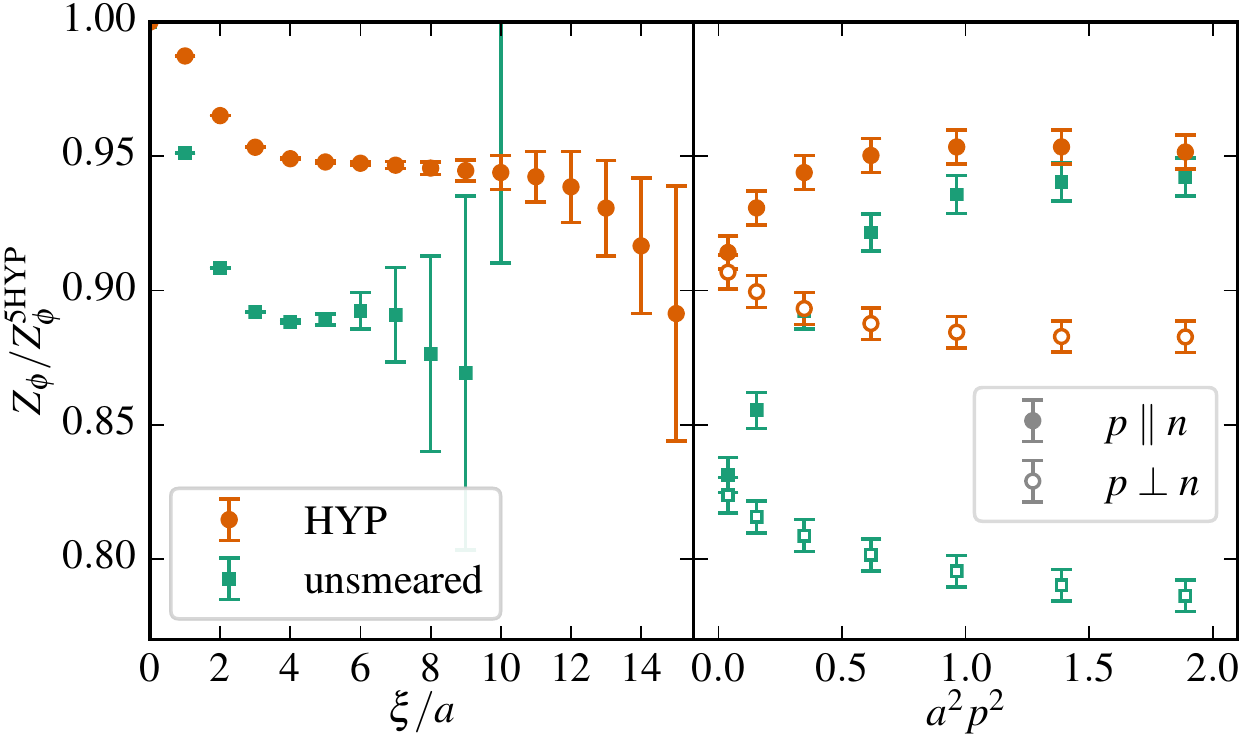}
  \caption{$Z_\phi$ for $\beta=2.10$, relative to the 5HYP case. Left:
    versus $\xi$, for $p\parallel n$ and $a^2p^2\approx 0.35$. Right:
    versus $p^2$, for $\xi=6a$ (unsmeared) and $\xi=10a$ (HYP). For
    each link discretization, we take the average of the two values at
    the smallest $p^2$.}
  \label{fig:Zratio_vs_both}
\end{figure}

\begin{figure}
  \centering
  \includegraphics[width=\columnwidth]{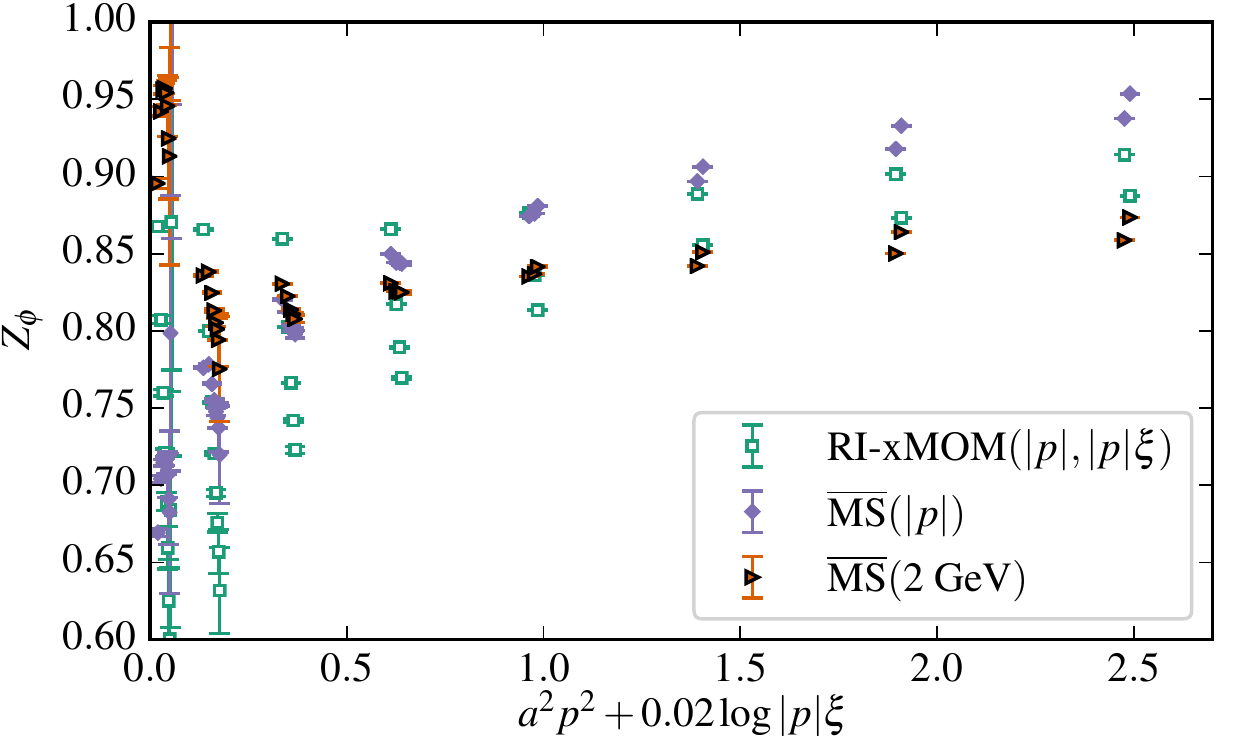}
  \caption{$Z_\phi$ for $\beta=2.10$, using unsmeared gauge links.
    Data are shown for a range of $p^2$ and $y\equiv |p|\xi$; the
    horizontal axis is $a^2p^2$, with a small displacement for
    different $y$ at the same $p^2$. The green open squares are given
    in our family of schemes, the blue filled diamonds show the result
    from conversion to $\MSbar$ at scale $|p|$ using
    Eq.~\eqref{eq:conversion}, and the orange filled triangles with
    black outlines show the $\MSbar$ results evolved to the scale
    2~GeV, using the two-loop anomalous dimension of the static-light
    current~\cite{Chetyrkin:2003vi,Ji:1991pr,Broadhurst:1991fz}.}
  \label{fig:Zphi}
\end{figure}

The remaining factor $Z_\phi$ depends on the kinematics that define
our scheme, but the ratio between $Z_\phi$ for different
discretizations is scheme independent. 
We evaluate this ratio in the plateau region at large $\xi$ and at
small $p$ (Fig.~\ref{fig:Zratio_vs_both}) to reduce lattice artifacts.
Finally, we determine $Z_\phi$ using unsmeared gauge links, converting
to the $\MSbar$ scheme and evolving to the scale 2~GeV, as shown in
Fig.~\ref{fig:Zphi}. We find that the one-loop conversion factor is
effective at removing much of the dependence on the scheme parameter
$|p|\xi$, and the two-loop evolution removes most of the dependence
on the scale $|p|$.

We apply these renormalization parameters to nucleon matrix elements
computed on one $N_f=2+1+1$ twisted mass ensemble at each lattice
spacing. The physical parameters are matched on the two ensembles:
$m_\pi\approx 370$~MeV and $p\cdot n\approx 1.85$~GeV. The coarser
ensemble was previously used in
Refs.~\cite{Alexandrou:2015rja,Alexandrou:2016jqi} and our methodology
is similar to Ref.~\cite{Alexandrou:2016jqi}, including the use of
momentum smearing~\cite{Bali:2016lva} in the nucleon interpolating
operator to obtain a good signal at large momentum.

\begin{figure*}
  \centering
  \includegraphics[width=0.495\textwidth]{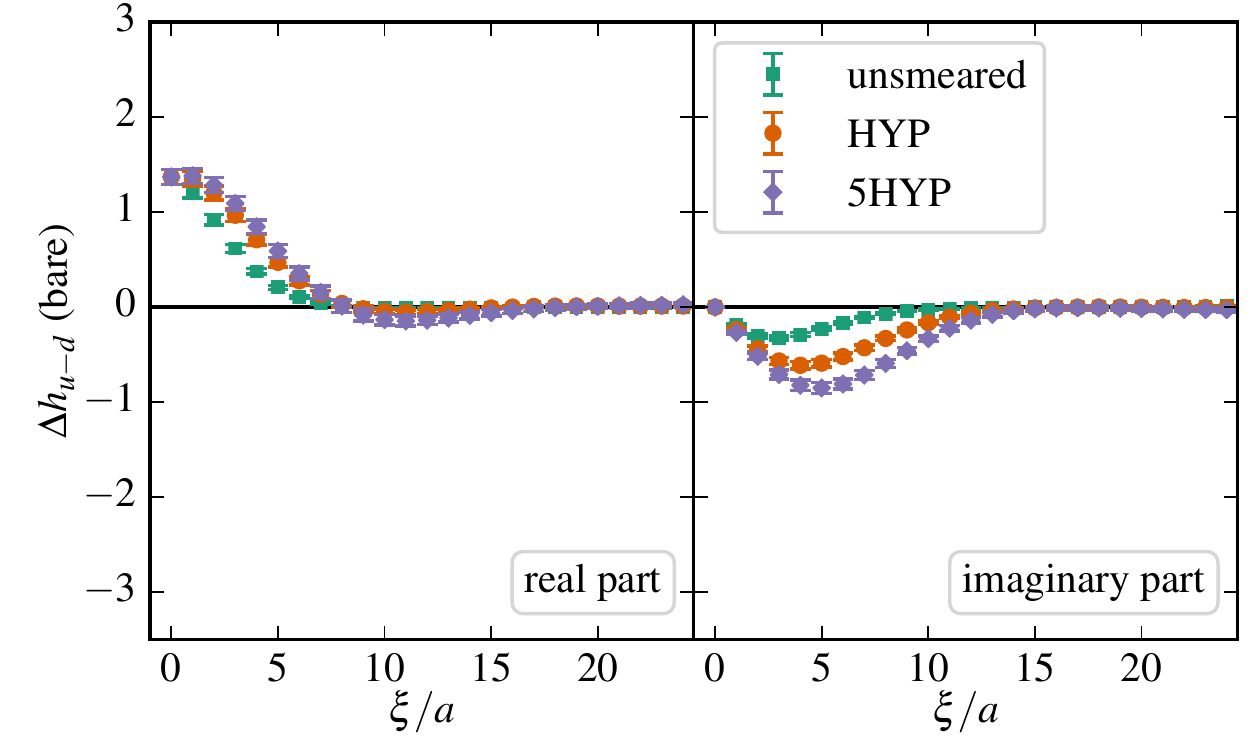}
  \includegraphics[width=0.495\textwidth]{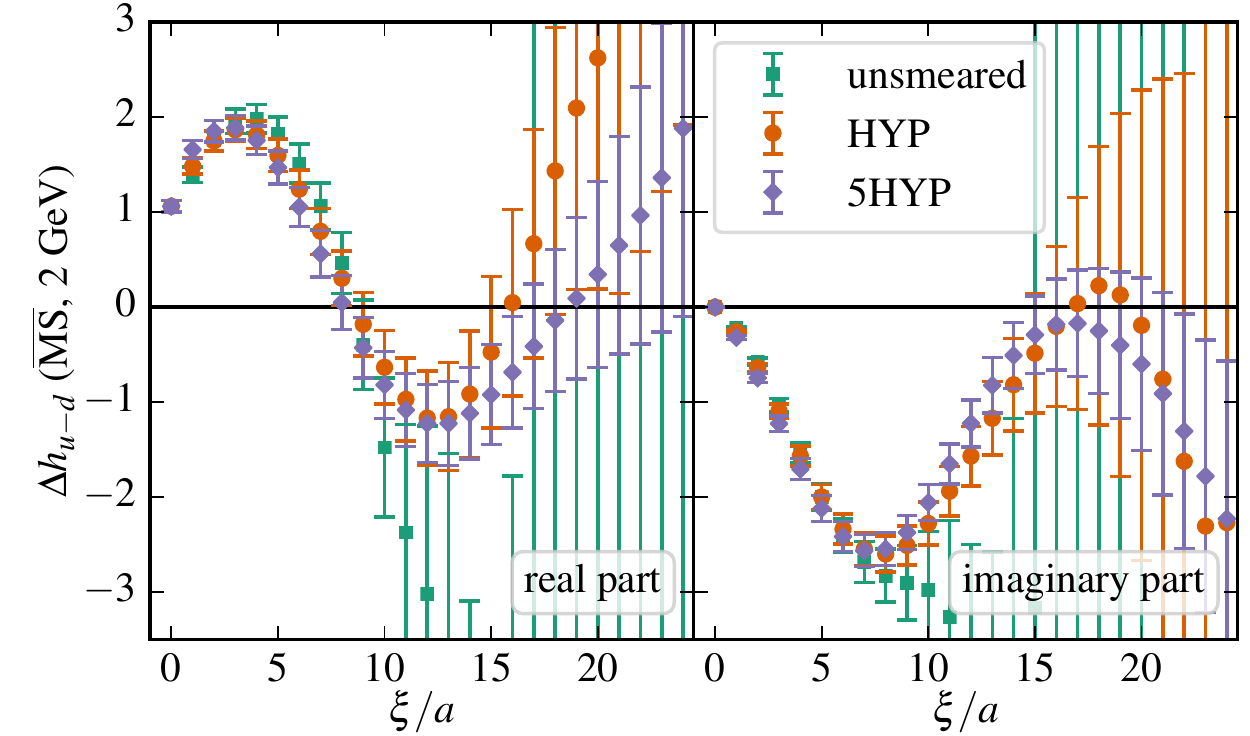}
  \caption{Matrix element for the helicity quasi-PDF versus $\xi$ on
    the $\beta=2.10$ ensemble, for three different link
    discretizations, bare (left) and renormalized (right). Only
    $\xi\geq 0$ is shown, since the real part is even in $\xi$ and the
    imaginary part is odd.}
  \label{fig:ff_smearing}
\end{figure*}

\begin{figure}
  \centering
  \includegraphics[width=\columnwidth]{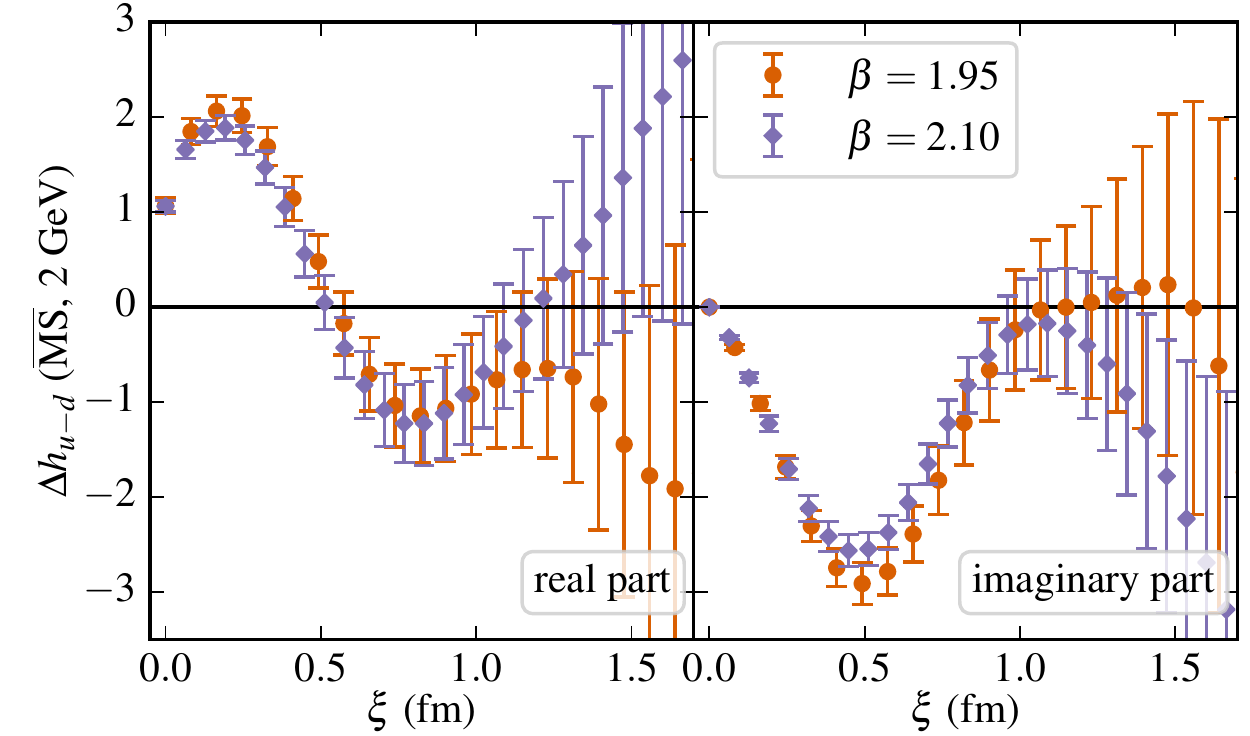}
  \caption{Renormalized matrix element for the helicity quasi-PDF
    versus $\xi$ on the two ensembles, using five steps of HYP
    smearing.}
  \label{fig:ff_ens}
\end{figure}

Figure~\ref{fig:ff_smearing} shows the effect of renormalization on
the isovector helicity matrix element $\Delta h_{u-d}$,
\begin{equation}
  \left\langle p,\lambda'\middle|\cO_{\slashed{n}\gamma_5}\middle|p,\lambda\right\rangle
\equiv\bar u(p,\lambda')\slashed{n}\gamma_5 u(p,\lambda) \Delta h(\xi,p\cdot n),
\end{equation}
computed on the fine ensemble, for different link
discretizations. Without renormalization there is a significant
disagreement between the different link types, and renormalization
brings them into good agreement. In the renormalized matrix elements,
we also see the benefit of smearing: without it, the statistical
errors grow rapidly at large $\xi$ and there is no useful signal for
$\xi\gtrsim 10a$. With smearing, we are able to see that the matrix
elements return toward zero at large $\xi$. Five steps of HYP smearing
also yields more precise data than one step, at large $\xi$. In
Fig.~\ref{fig:ff_ens}, we compare the renormalized data with five
steps of HYP smearing on the two ensembles. They are in excellent
agreement, which suggests that the linear divergence is under control
and discretization effects are not large.

\begin{figure}
  \centering
  \includegraphics[width=\columnwidth]{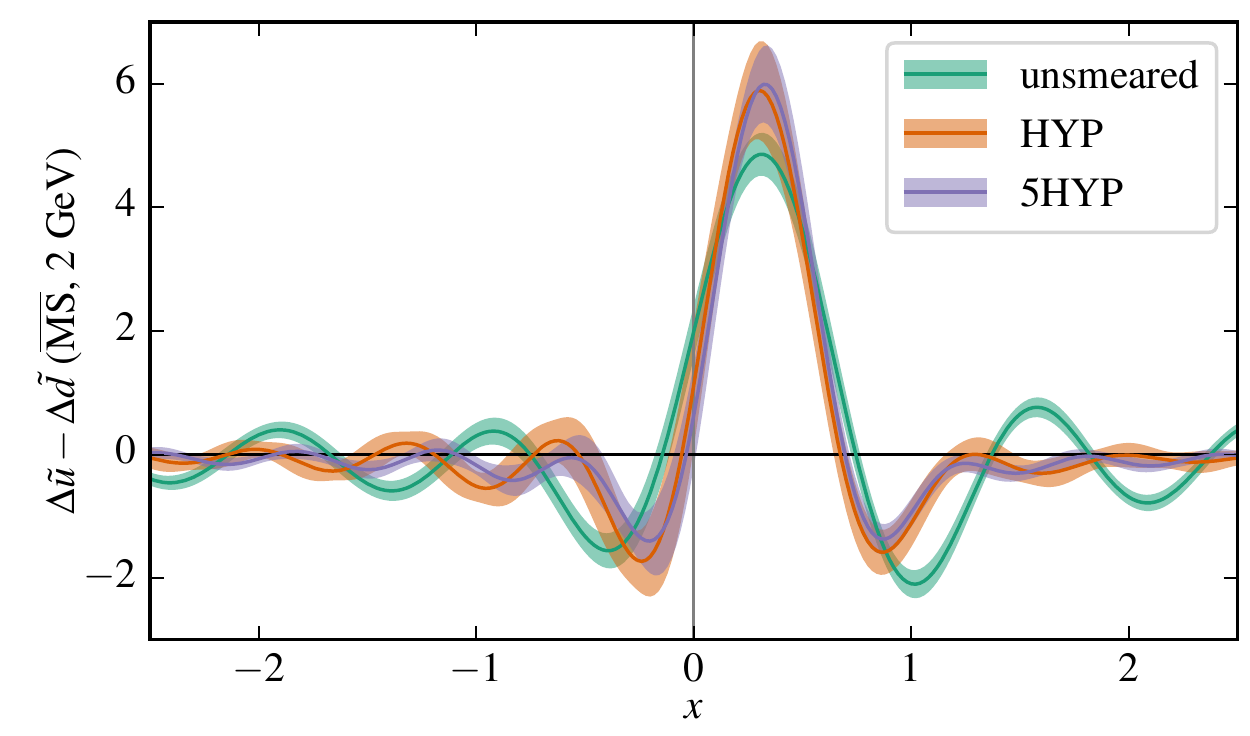}
  \caption{Isovector helicity quasi-PDF on the $\beta=2.10$ ensemble,
    for three different link discretizations, computed from
    renormalized matrix elements.}
  \label{fig:qpdf}
\end{figure}

Finally, we examine the helicity quasi-PDF, which is given by a
Fourier transform of the matrix element:
\begin{equation}
  \Delta\tilde q(x,p\cdot n) \equiv \frac{p\cdot n}{2\pi}\int d\xi\,
  e^{-ix\xi p\cdot n}\Delta h_q(\xi,p\cdot n).
\end{equation}
Without renormalization, unsmeared links lead to a much broader
distribution, as shown in Fig.~4 of Ref.~\cite{Alexandrou:2015rja}. We
show our renormalized results on the fine ensemble in
Fig.~\ref{fig:qpdf}. Because the data become noisy at large $\xi$, we
restrict the integral to $|\xi|\leq 16a$ ($|\xi|\leq 10a$ for the
unsmeared case). When this restriction removes part of
the signal, it leads to oscillations, which are clearly visible in the
unsmeared case. In future studies improved results could
be obtained by replacing the hard cutoff with a model for the
large-$\xi$ behavior of the matrix elements, or by applying one of the
methods proposed in Refs.~\cite{Lin:2017ani,Chen:2017lnm} for
suppressing the contributions at large $\xi$. Ignoring the
oscillations, we see that renormalization brings the data with
different link discretizations into reasonably good agreement.

In this work, we have shown that the nonlocal problem of renormalizing
lattice quasi-PDFs can be turned into a local problem by introducing
an auxiliary field. The auxiliary field approach can also be applied
to operators with staple-shaped gauge connection used for lattice
studies of transverse momentum-dependent
PDFs~\cite{Musch:2011er,Engelhardt:2015xja,Yoon:2017qzo}, where the
mixing pattern will be different. Because this approach is closely
connected with the static quark theory, it is possible to make use of
existing results from that theory such as the three-loop continuum
calculation in Ref.~\cite{Chetyrkin:2003vi}. We have contributed to
the evidence that link smearing is a very beneficial technique for
these calculations, as it leads to greatly reduced uncertainty at
large $\xi$; this was also explained some time ago in the static quark
theory~\cite{DellaMorte:2003mn}. There are several steps from the
results shown here to a full calculation of PDFs: matching from
quasi-PDFs to PDFs, control
over the $p\cdot n\to\infty$ limit, use of a physical pion mass, as
well as control over finite-volume and excited-state effects.

Note: after one of us presented a preliminary version of this work at
a conference~\cite{Green_Lat2017}, we were made aware of an
independent parallel effort based on the auxiliary field
approach~\cite{Ji:2017oey}. We also note another very recent article
discussing renormalizability of quasi-PDFs, without using the
auxiliary field approach~\cite{Ishikawa:2017faj}.

\begin{acknowledgments}
  We thank our colleagues in ETMC for a pleasant collaboration.
  Gauge fixing was performed using
  Fourier-accelerated conjugate gradient~\cite{Hudspith:2014oja},
  implemented in GLU~\cite{GLU}. Calculations were performed using the
  Grid library~\cite{Boyle:2016lbp} and the DD-$\alpha$AMG
  solver~\cite{Frommer:2013fsa} with twisted mass
  support~\cite{Alexandrou:2016izb}. The authors gratefully
  acknowledge the computing time granted by the John von Neumann
  Institute for Computing (NIC) and provided on the supercomputer
  JURECA~\cite{jureca} at Jülich Supercomputing Centre (JSC).
\end{acknowledgments}

\bibliography{aux_renorm}

\providecommand{\href}[2]{#2}\begingroup\raggedright\begin{thebibliography}{10}

\bibitem{Ji:2013dva}
X.~Ji, ``{Parton Physics on a Euclidean Lattice},''
  \href{http://dx.doi.org/10.1103/PhysRevLett.110.262002}{Phys. Rev. Lett.
  {\bfseries 110} (2013) 262002},
\href{http://arxiv.org/abs/1305.1539}{{\ttfamily arXiv:1305.1539 [hep-ph]}}.
%%CITATION = ARXIV:1305.1539;%%.

\bibitem{Xiong:2013bka}
X.~Xiong, X.~Ji, J.-H. Zhang, and Y.~Zhao, ``{One-loop matching for parton
  distributions: Nonsinglet case},''
  \href{http://dx.doi.org/10.1103/PhysRevD.90.014051}{Phys. Rev. D {\bfseries
  90} (2014) 014051},
\href{http://arxiv.org/abs/1310.7471}{{\ttfamily arXiv:1310.7471 [hep-ph]}}.
%%CITATION = ARXIV:1310.7471;%%.

\bibitem{Alexandrou:2015rja}
C.~Alexandrou, K.~Cichy, V.~Drach, E.~Garcia-Ramos, K.~Hadjiyiannakou,
  K.~Jansen, F.~Steffens, and C.~Wiese, ``{Lattice calculation of parton
  distributions},'' \href{http://dx.doi.org/10.1103/PhysRevD.92.014502}{Phys.
  Rev. D {\bfseries 92} (2015) 014502},
\href{http://arxiv.org/abs/1504.07455}{{\ttfamily arXiv:1504.07455 [hep-lat]}}.
%%CITATION = ARXIV:1504.07455;%%.

\bibitem{Ishikawa:2016znu}
T.~Ishikawa, Y.-Q. Ma, J.-W. Qiu, and S.~Yoshida, ``{Practical quasi parton
  distribution functions},''
\href{http://arxiv.org/abs/1609.02018}{{\ttfamily arXiv:1609.02018 [hep-lat]}}.
%%CITATION = ARXIV:1609.02018;%%.

\bibitem{Chen:2016fxx}
J.-W. Chen, X.~Ji, and J.-H. Zhang, ``{Improved quasi parton distribution
  through Wilson line renormalization},''
  \href{http://dx.doi.org/10.1016/j.nuclphysb.2016.12.004}{Nucl. Phys. B
  {\bfseries 915} (2017) 1--9},
\href{http://arxiv.org/abs/1609.08102}{{\ttfamily arXiv:1609.08102 [hep-ph]}}.
%%CITATION = ARXIV:1609.08102;%%.

\bibitem{Radyushkin:2017cyf}
A.~V. Radyushkin, ``{Quasi-parton distribution functions, momentum
  distributions, and pseudo-parton distribution functions},''
  \href{http://dx.doi.org/10.1103/PhysRevD.96.034025}{Phys. Rev. D {\bfseries
  96} (2017) 034025},
\href{http://arxiv.org/abs/1705.01488}{{\ttfamily arXiv:1705.01488 [hep-ph]}}.
%%CITATION = ARXIV:1705.01488;%%.

\bibitem{Orginos:2017kos}
K.~Orginos, A.~Radyushkin, J.~Karpie, and S.~Zafeiropoulos, ``{Lattice QCD
  exploration of parton pseudo-distribution functions},''
  \href{http://dx.doi.org/10.1103/PhysRevD.96.094503}{Phys. Rev. D {\bfseries
  96} (2017) 094503},
\href{http://arxiv.org/abs/1706.05373}{{\ttfamily arXiv:1706.05373 [hep-ph]}}.
%%CITATION = ARXIV:1706.05373;%%.

\bibitem{Musch:2011er}
B.~U. Musch, P.~Hägler, M.~Engelhardt, J.~W. Negele, and A.~Schäfer,
  ``{Sivers and Boer-Mulders observables from lattice QCD},''
  \href{http://dx.doi.org/10.1103/PhysRevD.85.094510}{Phys. Rev. D {\bfseries
  85} (2012) 094510},
\href{http://arxiv.org/abs/1111.4249}{{\ttfamily arXiv:1111.4249 [hep-lat]}}.
%%CITATION = ARXIV:1111.4249;%%.

\bibitem{Engelhardt:2015xja}
M.~Engelhardt, P.~Hägler, B.~Musch, J.~Negele, and A.~Schäfer, ``{Lattice QCD
  study of the Boer-Mulders effect in a pion},''
  \href{http://dx.doi.org/10.1103/PhysRevD.93.054501}{Phys. Rev. D {\bfseries
  93} (2016) 054501},
\href{http://arxiv.org/abs/1506.07826}{{\ttfamily arXiv:1506.07826 [hep-lat]}}.
%%CITATION = ARXIV:1506.07826;%%.

\bibitem{Yoon:2017qzo}
B.~Yoon, M.~Engelhardt, R.~Gupta, T.~Bhattacharya, J.~R. Green, B.~U. Musch,
  J.~W. Negele, A.~V. Pochinsky, A.~Schäfer, and S.~N. Syritsyn, ``Nucleon
  transverse momentum-dependent parton distributions in lattice {QCD}:
  {R}enormalization patterns and discretization effects,''
  \href{http://dx.doi.org/10.1103/PhysRevD.96.094508}{Phys. Rev. D {\bfseries
  96} (2017) 094508},
\href{http://arxiv.org/abs/1706.03406}{{\ttfamily arXiv:1706.03406 [hep-lat]}}.
%%CITATION = ARXIV:1706.03406;%%.

\bibitem{Lin:2014zya}
H.-W. Lin, J.-W. Chen, S.~D. Cohen, and X.~Ji, ``{Flavor Structure of the
  Nucleon Sea from Lattice QCD},''
  \href{http://dx.doi.org/10.1103/PhysRevD.91.054510}{Phys. Rev. D {\bfseries
  91} (2015) 054510},
\href{http://arxiv.org/abs/1402.1462}{{\ttfamily arXiv:1402.1462 [hep-ph]}}.
%%CITATION = ARXIV:1402.1462;%%.

\bibitem{Chen:2016utp}
J.-W. Chen, S.~D. Cohen, X.~Ji, H.-W. Lin, and J.-H. Zhang, ``{Nucleon Helicity
  and Transversity Parton Distributions from Lattice QCD},''
  \href{http://dx.doi.org/10.1016/j.nuclphysb.2016.07.033}{Nucl. Phys. B
  {\bfseries 911} (2016) 246--273},
\href{http://arxiv.org/abs/1603.06664}{{\ttfamily arXiv:1603.06664 [hep-ph]}}.
%%CITATION = ARXIV:1603.06664;%%.

\bibitem{Alexandrou:2016jqi}
C.~Alexandrou, K.~Cichy, M.~Constantinou, K.~Hadjiyiannakou, K.~Jansen,
  F.~Steffens, and C.~Wiese, ``Updated lattice results for parton
  distributions,'' \href{http://dx.doi.org/10.1103/PhysRevD.96.014513}{Phys.
  Rev. D {\bfseries 96} (2017) 014513},
\href{http://arxiv.org/abs/1610.03689}{{\ttfamily arXiv:1610.03689 [hep-lat]}}.
%%CITATION = ARXIV:1610.03689;%%.

\bibitem{Constantinou:2017sej}
M.~Constantinou and H.~Panagopoulos, ``{Perturbative renormalization of
  quasi-parton distribution functions},''
  \href{http://dx.doi.org/10.1103/PhysRevD.96.054506}{Phys. Rev. D {\bfseries
  96} (2017) 054506},
\href{http://arxiv.org/abs/1705.11193}{{\ttfamily arXiv:1705.11193 [hep-lat]}}.
%%CITATION = ARXIV:1705.11193;%%.

\bibitem{Alexandrou:2017huk}
C.~Alexandrou, K.~Cichy, M.~Constantinou, K.~Hadjiyiannakou, K.~Jansen,
  H.~Panagopoulos, and F.~Steffens, ``{A complete non-perturbative
  renormalization prescription for quasi-PDFs},''
  \href{http://dx.doi.org/10.1016/j.nuclphysb.2017.08.012}{Nucl. Phys. B
  {\bfseries 923} (2017) 394--415},
\href{http://arxiv.org/abs/1706.00265}{{\ttfamily arXiv:1706.00265 [hep-lat]}}.
%%CITATION = ARXIV:1706.00265;%%.

\bibitem{Chen:2017mzz}
J.-W. Chen, T.~Ishikawa, L.~Jin, H.-W. Lin, Y.-B. Yang, J.-H. Zhang, and
  Y.~Zhao, ``Parton distribution function with nonperturbative renormalization
  from lattice {QCD},''
  \href{http://dx.doi.org/10.1103/PhysRevD.97.014505}{Phys. Rev. D {\bfseries
  97} (2018) 014505},
\href{http://arxiv.org/abs/1706.01295}{{\ttfamily arXiv:1706.01295 [hep-lat]}}.
%%CITATION = ARXIV:1706.01295;%%.

\bibitem{Martinelli:1994ty}
G.~Martinelli, C.~Pittori, C.~T. Sachrajda, M.~Testa, and A.~Vladikas, ``{A
  General method for nonperturbative renormalization of lattice operators},''
  \href{http://dx.doi.org/10.1016/0550-3213(95)00126-D}{Nucl. Phys. B
  {\bfseries 445} (1995) 81--108},
\href{http://arxiv.org/abs/hep-lat/9411010}{{\ttfamily arXiv:hep-lat/9411010}}.
%%CITATION = HEP-LAT/9411010;%%.

\bibitem{Craigie:1980qs}
N.~S. Craigie and H.~Dorn, ``{On the Renormalization and Short Distance
  Properties of Hadronic Operators in {QCD}},''
\href{http://dx.doi.org/10.1016/0550-3213(81)90372-2}{Nucl. Phys. B {\bfseries
  185} (1981) 204--220}.
%%CITATION = NUPHA,B185,204;%%.

\bibitem{Dorn:1986dt}
H.~Dorn, ``{Renormalization of Path Ordered Phase Factors and Related Hadron
  Operators in Gauge Field Theories},''
\href{http://dx.doi.org/10.1002/prop.19860340104}{Fortsch. Phys. {\bfseries 34}
  (1986) 11--56}.
%%CITATION = FPYKA,34,11;%%.

\bibitem{Constantinou_privcomm}
M.~Constantinou. Private communication.

\bibitem{Alexandrou:2015sea}
{\bfseries ETM} Collaboration, C.~Alexandrou, M.~Constantinou, and
  H.~Panagopoulos, ``{Renormalization functions for $N_f=2$ and $N_f=4$ twisted
  mass fermions},'' \href{http://dx.doi.org/10.1103/PhysRevD.95.034505}{Phys.
  Rev. D {\bfseries 95} (2017) 034505},
\href{http://arxiv.org/abs/1509.00213}{{\ttfamily arXiv:1509.00213 [hep-lat]}}.
%%CITATION = ARXIV:1509.00213;%%.

\bibitem{Eichten:1989zv}
E.~Eichten and B.~R. Hill, ``{An Effective Field Theory for the Calculation of
  Matrix Elements Involving Heavy Quarks},''
\href{http://dx.doi.org/10.1016/0370-2693(90)92049-O}{Phys. Lett. B {\bfseries
  234} (1990) 511--516}.
%%CITATION = PHLTA,B234,511;%%.

\bibitem{Sommer:2010ic}
R.~Sommer,
  \href{http://dx.doi.org/10.1093/acprof:oso/9780199691609.003.0009}{``{Introduction
  to Non-perturbative Heavy Quark Effective Theory},''} in {\em {Modern
  perspectives in lattice QCD: Quantum field theory and high performance
  computing. Proceedings, International School, 93rd Session, Les Houches,
  France, August 3-28, 2009}}, pp.~517--590.
\newblock 2010.
\newblock
\href{http://arxiv.org/abs/1008.0710}{{\ttfamily arXiv:1008.0710 [hep-lat]}}.
\newblock
%%CITATION = ARXIV:1008.0710;%%.

\bibitem{Kurth:2000ki}
{\bfseries ALPHA} Collaboration, M.~Kurth and R.~Sommer, ``{Renormalization and
  $O(a)$ improvement of the static axial current},''
  \href{http://dx.doi.org/10.1016/S0550-3213(00)00750-1}{Nucl. Phys. B
  {\bfseries 597} (2001) 488--518},
\href{http://arxiv.org/abs/hep-lat/0007002}{{\ttfamily arXiv:hep-lat/0007002}}.
%%CITATION = HEP-LAT/0007002;%%.

\bibitem{Ji:2015jwa}
X.~Ji and J.-H. Zhang, ``{Renormalization of quasiparton distribution},''
  \href{http://dx.doi.org/10.1103/PhysRevD.92.034006}{Phys. Rev. D {\bfseries
  92} (2015) 034006},
\href{http://arxiv.org/abs/1505.07699}{{\ttfamily arXiv:1505.07699 [hep-ph]}}.
%%CITATION = ARXIV:1505.07699;%%.

\bibitem{Sturm:2009kb}
C.~Sturm, Y.~Aoki, N.~H. Christ, T.~Izubuchi, C.~T.~C. Sachrajda, and A.~Soni,
  ``{Renormalization of quark bilinear operators in a momentum-subtraction
  scheme with a nonexceptional subtraction point},''
  \href{http://dx.doi.org/10.1103/PhysRevD.80.014501}{Phys. Rev. D {\bfseries
  80} (2009) 014501},
\href{http://arxiv.org/abs/0901.2599}{{\ttfamily arXiv:0901.2599 [hep-ph]}}.
%%CITATION = ARXIV:0901.2599;%%.

\bibitem{Green_forthcoming}
J.~Green, K.~Jansen, and F.~Steffens. Forthcoming.

\bibitem{Chetyrkin:2003vi}
K.~G. Chetyrkin and A.~G. Grozin, ``{Three loop anomalous dimension of the
  heavy light quark current in HQET},''
  \href{http://dx.doi.org/10.1016/S0550-3213(03)00490-5}{Nucl. Phys. B
  {\bfseries 666} (2003) 289--302},
\href{http://arxiv.org/abs/hep-ph/0303113}{{\ttfamily arXiv:hep-ph/0303113}}.
%%CITATION = HEP-PH/0303113;%%.

\bibitem{Melnikov:2000zc}
K.~Melnikov and T.~van Ritbergen, ``The three loop on-shell renormalization of
  {QCD and QED},'' \href{http://dx.doi.org/10.1016/S0550-3213(00)00526-5}{Nucl.
  Phys. B {\bfseries 591} (2000) 515--546},
\href{http://arxiv.org/abs/hep-ph/0005131}{{\ttfamily arXiv:hep-ph/0005131}}.
%%CITATION = HEP-PH/0005131;%%.

\bibitem{Musch:2010ka}
B.~U. Musch, P.~Hägler, J.~W. Negele, and A.~Schäfer, ``{Exploring quark
  transverse momentum distributions with lattice QCD},''
  \href{http://dx.doi.org/10.1103/PhysRevD.83.094507}{Phys. Rev. D {\bfseries
  83} (2011) 094507},
\href{http://arxiv.org/abs/1011.1213}{{\ttfamily arXiv:1011.1213 [hep-lat]}}.
%%CITATION = ARXIV:1011.1213;%%.

\bibitem{Hasenfratz:2001hp}
A.~Hasenfratz and F.~Knechtli, ``{Flavor symmetry and the static potential with
  hypercubic blocking},''
  \href{http://dx.doi.org/10.1103/PhysRevD.64.034504}{Phys. Rev. D {\bfseries
  64} (2001) 034504},
\href{http://arxiv.org/abs/hep-lat/0103029}{{\ttfamily arXiv:hep-lat/0103029}}.
%%CITATION = HEP-LAT/0103029;%%.

\bibitem{Frezzotti:2003ni}
R.~Frezzotti and G.~C. Rossi, ``{Chirally improving Wilson fermions 1. $O(a)$
  improvement},'' \href{http://dx.doi.org/10.1088/1126-6708/2004/08/007}{JHEP
  {\bfseries 08} (2004) 007},
\href{http://arxiv.org/abs/hep-lat/0306014}{{\ttfamily arXiv:hep-lat/0306014}}.
%%CITATION = HEP-LAT/0306014;%%.

\bibitem{Ji:1991pr}
X.~Ji and M.~J. Musolf, ``{Sub-leading logarithmic mass-dependence in
  heavy-meson form-factors},''
\href{http://dx.doi.org/10.1016/0370-2693(91)91916-J}{Phys. Lett. B {\bfseries
  257} (1991) 409--413}.
%%CITATION = PHLTA,B257,409;%%.

\bibitem{Broadhurst:1991fz}
D.~J. Broadhurst and A.~G. Grozin, ``{Two-loop renormalization of the effective
  field theory of a static quark},''
  \href{http://dx.doi.org/10.1016/0370-2693(91)90532-U}{Phys. Lett. B
  {\bfseries 267} (1991) 105--110},
\href{http://arxiv.org/abs/hep-ph/9908362}{{\ttfamily arXiv:hep-ph/9908362}}.
%%CITATION = HEP-PH/9908362;%%.

\bibitem{Bali:2016lva}
G.~S. Bali, B.~Lang, B.~U. Musch, and A.~Schäfer, ``{Novel quark smearing for
  hadrons with high momenta in lattice QCD},''
  \href{http://dx.doi.org/10.1103/PhysRevD.93.094515}{Phys. Rev. D {\bfseries
  93} (2016) 094515},
\href{http://arxiv.org/abs/1602.05525}{{\ttfamily arXiv:1602.05525 [hep-lat]}}.
%%CITATION = ARXIV:1602.05525;%%.

\bibitem{Lin:2017ani}
H.-W. Lin, J.-W. Chen, T.~Ishikawa, and J.-H. Zhang, ``{Improved Parton
  Distribution Functions at Physical Pion Mass},''
\href{http://arxiv.org/abs/1708.05301}{{\ttfamily arXiv:1708.05301 [hep-lat]}}.
%%CITATION = ARXIV:1708.05301;%%.

\bibitem{Chen:2017lnm}
J.-W. Chen, T.~Ishikawa, L.~Jin, H.-W. Lin, A.~Schäfer, Y.-B. Yang, J.-H.
  Zhang, and Y.~Zhao, ``{Gaussian-weighted Parton Quasi-distribution},''
\href{http://arxiv.org/abs/1711.07858}{{\ttfamily arXiv:1711.07858 [hep-ph]}}.
%%CITATION = ARXIV:1711.07858;%%.

\bibitem{DellaMorte:2003mn}
{\bfseries ALPHA} Collaboration, M.~Della~Morte, S.~Dürr, J.~Heitger,
  H.~Molke, J.~Rolf, A.~Shindler, and R.~Sommer, ``{Lattice HQET with
  exponentially improved statistical precision},''
  \href{http://dx.doi.org/10.1016/j.physletb.2005.03.017}{Phys. Lett. B
  {\bfseries 581} (2004) 93--98},
  \href{http://arxiv.org/abs/hep-lat/0307021}{{\ttfamily
  arXiv:hep-lat/0307021}}.
[Erratum: Phys. Lett. B 612 (2005) 313].
%%CITATION = HEP-LAT/0307021;%%.

\bibitem{Green_Lat2017}
J.~Green, ``{Auxiliary field approach to extended operators for quasi-PDFs}.''
  {Talk presented at the 35th International Symposium on Lattice Field Theory
  (Lattice 2017), Granada, Spain}, June, 2017.
\newblock \url{https://makondo.ugr.es/event/0/session/95/contribution/332}.

\bibitem{Ji:2017oey}
X.~Ji, J.-H. Zhang, and Y.~Zhao, ``{Renormalization in Large Momentum Effective
  Theory of Parton Physics},''
\href{http://arxiv.org/abs/1706.08962}{{\ttfamily arXiv:1706.08962 [hep-ph]}}.
%%CITATION = ARXIV:1706.08962;%%.

\bibitem{Ishikawa:2017faj}
T.~Ishikawa, Y.-Q. Ma, J.-W. Qiu, and S.~Yoshida, ``{Renormalizability of
  quasiparton distribution functions},''
  \href{http://dx.doi.org/10.1103/PhysRevD.96.094019}{Phys. Rev. D {\bfseries
  96} (2017) 094019},
\href{http://arxiv.org/abs/1707.03107}{{\ttfamily arXiv:1707.03107 [hep-ph]}}.
%%CITATION = ARXIV:1707.03107;%%.

\bibitem{Hudspith:2014oja}
{\bfseries RBC, UKQCD} Collaboration, R.~J. Hudspith, ``{Fourier Accelerated
  Conjugate Gradient Lattice Gauge Fixing},''
  \href{http://dx.doi.org/10.1016/j.cpc.2014.10.017}{Comput. Phys. Commun.
  {\bfseries 187} (2015) 115--119},
\href{http://arxiv.org/abs/1405.5812}{{\ttfamily arXiv:1405.5812 [hep-lat]}}.
%%CITATION = ARXIV:1405.5812;%%.

\bibitem{GLU}
R.~J. Hudspith, ``{Gauge Link Utility}.''
  {{\href{https://github.com/RJhudspith/GLU}{\texttt{https://github.com/RJhudspith/GLU}}}}.

\bibitem{Boyle:2016lbp}
P.~A. Boyle, G.~Cossu, A.~Yamaguchi, and A.~Portelli, ``{Grid: A next
  generation data parallel C++ QCD library},'' PoS {\bfseries LATTICE2015}
  (2016) 023,
\href{http://arxiv.org/abs/1512.03487}{{\ttfamily arXiv:1512.03487 [hep-lat]}}.
%%CITATION = POSCI,LATTICE2015,023;%%.

\bibitem{Frommer:2013fsa}
A.~Frommer, K.~Kahl, S.~Krieg, B.~Leder, and M.~Rottmann, ``{Adaptive
  Aggregation-Based Domain Decomposition Multigrid for the Lattice
  Wilson--Dirac Operator},'' \href{http://dx.doi.org/10.1137/130919507}{SIAM J.
  Sci. Comput. {\bfseries 36} (2014) A1581--A1608},
\href{http://arxiv.org/abs/1303.1377}{{\ttfamily arXiv:1303.1377 [hep-lat]}}.
%%CITATION = ARXIV:1303.1377;%%.

\bibitem{Alexandrou:2016izb}
C.~Alexandrou, S.~Bacchio, J.~Finkenrath, A.~Frommer, K.~Kahl, and M.~Rottmann,
  ``{Adaptive Aggregation-based Domain Decomposition Multigrid for Twisted Mass
  Fermions},'' \href{http://dx.doi.org/10.1103/PhysRevD.94.114509}{Phys. Rev. D
  {\bfseries 94} (2016) 114509},
\href{http://arxiv.org/abs/1610.02370}{{\ttfamily arXiv:1610.02370 [hep-lat]}}.
%%CITATION = ARXIV:1610.02370;%%.

\bibitem{jureca}
{Jülich Supercomputing Centre}, ``{JURECA: General-purpose supercomputer at
  Jülich Supercomputing Centre},''
  \href{http://dx.doi.org/10.17815/jlsrf-2-121}{Journal of large-scale research
  facilities {\bfseries 2} (2016) A62}.

\end{thebibliography}\endgroup
\bibliographystyle{utphys-noitalics}

\end{document}